\documentclass{aa}
\usepackage{graphicx}

\newcommand{\corot}{{\textsc{CoRoT}}}
\newcommand{\cible}{{HD\,175726}}

\newcommand{\ind}[1]{_{\mathrm{#1}}}

\def\muHz{\,$\mu$Hz}         
\def\mHz{\,mHz}
\def\m2s2{\,m$^{2}$\,s$^{-2}$} 
\def\kms{\,km\,s$^{-1}$}       
\def\vsini{$v\sin i$}          

\def\tauto{t\ind{m}}
\def\nucen{\nu\ind{w,c}}
\def\window{\delta\nu\ind{w,e}}
\def\deltanunu{\Delta\nu(\nu)}

\begin{document}
\title{
The CoRoT target HD\,175726:\\ an active star with weak solar-like oscillations.
\thanks{The CoRoT space mission, launched on 2006 December 27, was developed and is operated by the CNES, with participation of the Science Programs of ESA, ESA's RSSD, Austria, Belgium, Brazil,
Germany and Spain.}}
\titlerunning{Asteroseismic study of HD 175726}
\author{
B. Mosser\inst{1}\and
E. Michel \inst{1}\and
T. Appourchaux\inst{2}\and
C. Barban\inst{1}\and
F. Baudin\inst{2}\and
P. Boumier\inst{2}\and
H. Bruntt\inst{1,3}\and
C. Catala\inst{1}\and
S. Deheuvels \inst{1}\and
R.A. Garcia\inst{4}\and
P. Gaulme\inst{2}\and
C. Regulo\inst{5,6}\and
I. Roxburgh\inst{7,1}\and
R. Samadi \inst{1}\and
G. Verner\inst{7}\and
M. Auvergne\inst{1}\and
A. Baglin\inst{1}\and
J. Ballot\inst{8}\and
O. Benomar\inst{2}\and
S. Mathur\inst{9}}

\offprints{B. Mosser}

\institute{LESIA, CNRS, Universit\'e Pierre et Marie Curie, Universit\'e Denis Diderot, Observatoire de Paris, 92195 Meudon cedex, France\\
\email{benoit.mosser@obspm.fr}
\and
Institut d'Astrophysique Spatiale, UMR8617, Universit\'e Paris XI, B\^atiment 121, 91405 Orsay Cedex, France
\and
Sydney Institute for Astronomy, School of Physics, The University of Sydney, NSW 2006, Australia
\and
Laboratoire AIM, CEA/DSM-CNRS - Univ. Paris 7 Diderot - IRFU/SAp, F-91191 Gif-sur-Yvette Cedex, France
\and
Instituto de Astrof\'isica de Canarias, 38205 La Laguna, Tenerife, Spain
\and
Universidad de la Laguna, 38206 La Laguna, Tenerife, Spain
\and
Astronomy Unit, Queen Mary, University of London Mile End Road, London E1 4NS, UK
\and
Laboratoire d'Astrophysique de Toulouse-Tarbes, Universit\'e de Toulouse,
CNRS, 14 av. Edouard Belin, F-31400 Toulouse, France
\and
Indian Institute of Astrophysics, Bangalore, India
}
\date{Received 23 February 2009 / Accepted 5 July 2009}

\abstract{The CoRoT short runs give us the opportunity to observe a large variety of late-type stars through their solar-like oscillations. We report observations of the star HD\,175726 that lasted for 27 days during the first short run of the mission. The time series reveals a high-activity signal and the power spectrum presents an excess due to solar-like oscillations with a low signal-to-noise ratio.}%
{Our aim is to identify the most efficient tools to extract as much information as possible from the power density spectrum.}%
{The most productive method appears to be the autocorrelation of the time series, calculated as the spectrum of the filtered spectrum. This method is efficient, very rapid computationally, and will be useful for the analysis of other targets, observed with CoRoT or with forthcoming missions such as Kepler and Plato.}%
{The mean large separation has been measured to be $97.2\pm0.5$\muHz, slightly below the expected value determined from solar scaling laws. We also show strong evidence for variation of the large separation with frequency. The bolometric mode amplitude is only $1.7\pm0.25$\,ppm for radial modes, which is 1.7 times less than expected. Due to the low signal-to-noise ratio, mode identification is not possible for the available data set of HD\,175726.}%
{This study shows the possibility of extracting a seismic signal despite a signal-to-noise ratio of only 0.37. The observation of such a target shows the efficiency of the CoRoT data, and the potential benefit of longer observing runs.}

\maketitle
\section{Introduction}

CoRoT (COnvection, ROtation and planetary Transits) is a satellite developed by the French space agency (Centre National d'\'Etudes Spatiales, CNES), with participation of the Science Program of ESA, Austria, Belgium, Brazil, Germany and Spain. The scientific objectives of CoRoT are to detect exoplanets and to study the interiors of stars thanks to its high-performance photometric observations.
The program of the CoRoT mission provides short runs in between 5-month long runs. Such short runs allow us to study a larger set of variable stars. \cible, a solar-like star suspected to show measurable solar-like oscillations, was the main target of the first short run, and was observed for 27 days in October 2007.

In Section~\ref{etoile}, we discuss the physical parameters of \cible\ and the prediction of the asteroseismic signal by scaling from the Sun. Observations are presented in Sect.~\ref{observations}, with the analysis of the time series and of the activity signal. The analysis of the power spectrum is analysed in Sect.~\ref{signature}, with the identification of the large separation and of its variation with frequency. Section~\ref{discussion} is devoted to discussions, Section~\ref{conclusion} to conclusions.


\section{Stellar parameters\label{etoile}}

\cible, or HIP 92984, is known as an F9/G0 dwarf. It is the third component of $\theta$ Serpentis, separated by 7' from the two major components. Its primary parameters  were given by  \cite{2004A&A...418..989N}, using Str\"omgren photometry, with values revisited in \cite{2007A&A...475..519H}. \cible\ belongs to their large kinematically unbiased sample of $\simeq$16\,000 nearby F and G dwarf stars, for which metallicity, rotation, age, kinematics, and galactic orbits were determined. During the preparation of the mission, this star was also the scope of a refined analysis by \cite{2006A&A...448..341G}. The temperature and $\log g$ value in \cite{2006A&A...448..341G} are significantly different from other determinations. Their parameters have not been taken into account for our estimate of the stellar asteroseismic parameters, since they seem incompatible with other values and indicates a star that is too dense. \cite{bruntt2009} proposed new parameters that take into account very high signal-to-noise ratio high-resolution spectrometric measurements.  Concerning rotation, \cite{2004A&A...427..933K} give a measurement of the projected rotational velocity of about 13.5\kms, whereas \cite{2002A&A...384..491C} give 12\kms. The values and uncertainties of the primary parameters proposed by \cite{bruntt2009} are reported in Table~\ref{prop-phys}.

The estimated mass and radius derived from the physical parameters are very close to the solar values.
We therefore expect the value of the large separation to be close to the solar value too, namely around 132$\pm$10\muHz, according to the 1-$\sigma$ error bars (\cite{1995A&A...293...87K}). We nevertheless have to take into account that the estimated mass and radius depend on the physical modeling of the stellar photosphere. Furthermore, experience tells us that we have to consider 2-$\sigma$ error bars for a reliable prediction. Therefore, we should search for a large separation in the range 132$\pm$20\muHz.

Similarly derived from scaling laws, the maximum amplitude should occur around 2.9$\pm$0.7\,mHz  (\cite{1995A&A...293...87K}). Also, from the $(L/M)^{0.7}$ amplitude law reported from Doppler measurements by \cite{2007A&A...463..297S} and adapted for photometric measurement by \cite{2008Sci...322..558M}, we should expect a maximum amplitude of about 2.9$\pm$0.3\,ppm for \cible. All these inferred values are used to delimit the search for the asteroseismic signal.

\begin{table}
\caption{Primary parameters of \cible}\label{prop-phys}.
\centering
\begin{tabular}{lr}
\hline
$T\ind{eff} (K)$   & 6040$\pm$80\\
$[$Fe/H$]\ind{f}$     & $-0.10\pm0.10$\\
$m\ind{V}$    & 6.72\\
$M\ind{V}$    & 4.56\\
$L/L_\odot$   & 1.21$\pm$0.07\\
$\Pi$   (mas)  & 37.00$\pm$0.84\\
\hline
$\log g$ (cm\,s$^{-2}$)    & 4.44$\pm$0.10  \\
$v \sin i$ (\kms) & 13.5$\pm$0.5\\
\hline
$M/M_\odot$   & 0.993$\pm$0.060 \\
$R/R_\odot$   & 1.014$\pm$0.035 \\
\hline
\end{tabular}

Reference: \cite{bruntt2009}.
\end{table}

\begin{figure}
\centering
\includegraphics[width=8.5cm]{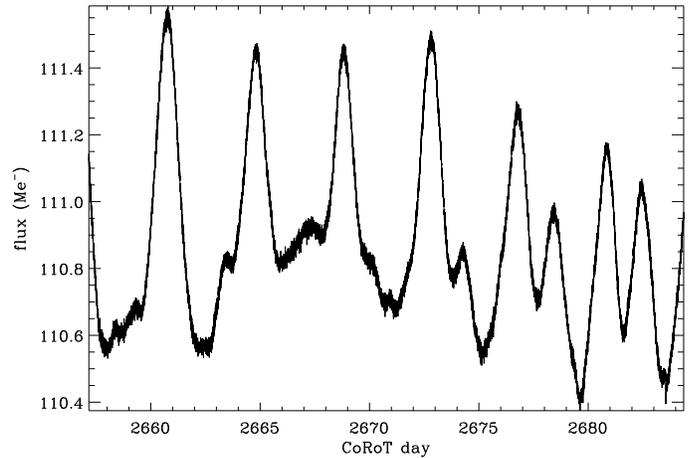}
\vskip 0.2cm
\caption{Level-2 light curve of \cible. The maximum variations are typically 1\,\% peak-to-peak. }
\label{intensity}
\end{figure}

\begin{figure}
\centering
\includegraphics[width=8.5cm]{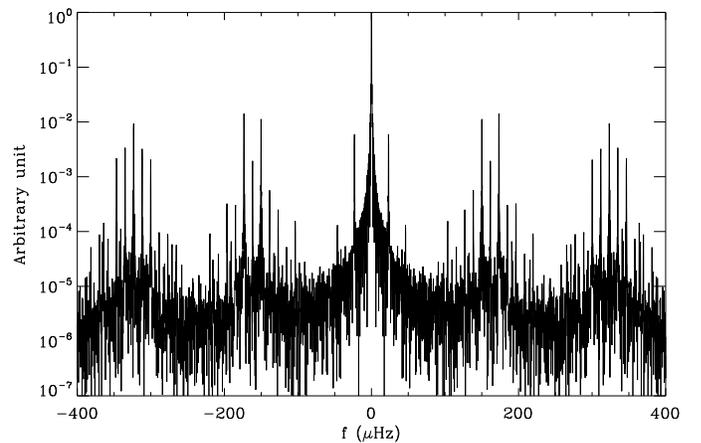}
\caption{Power spectrum of the window. Side lobes occur at the harmonics of the orbital frequency, modulated by the diurnal frequency.
\label{specfen}}
\end{figure}

\begin{figure}
\centering
\includegraphics[width=8.5cm]{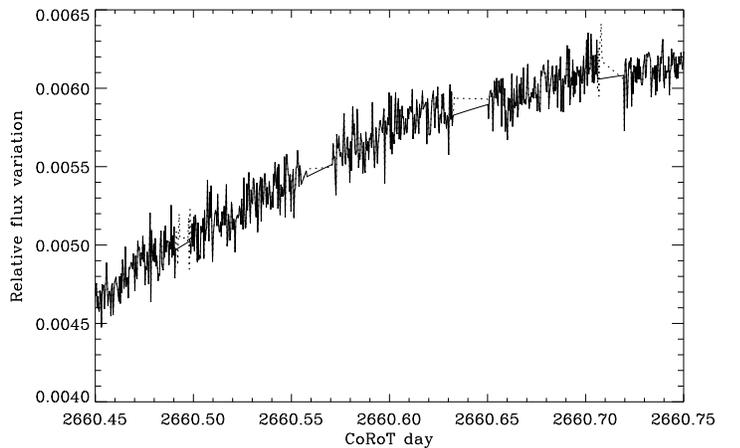}
\caption{Sample of the time series, with the original data (dotted line) and the corrected data based on an interpolated time series for filling the gaps (solid line).
\label{interpol}}
\end{figure}

\begin{figure}
\centering
\includegraphics[width=8.5cm]{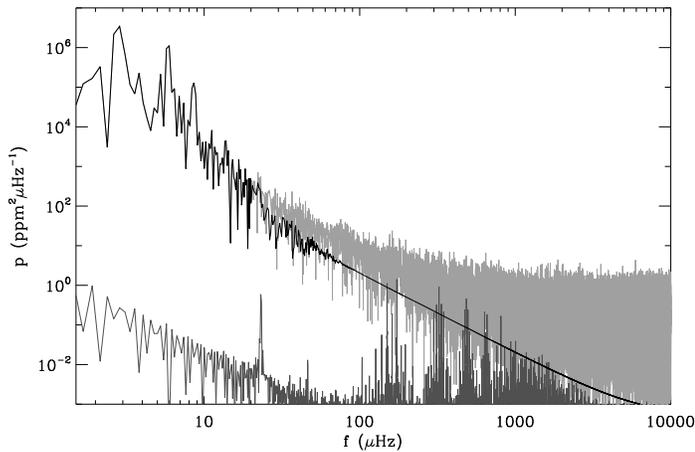}
\caption{Power density spectra of the corrected time series (light grey line), of the low-frequency interpolation (black line) and, with an adapted scale, of the window (dark grey line). Note the axes use logarithmic scale.
\label{specbf}}
\end{figure}

\begin{figure}
\centering
\includegraphics[width=8.5cm]{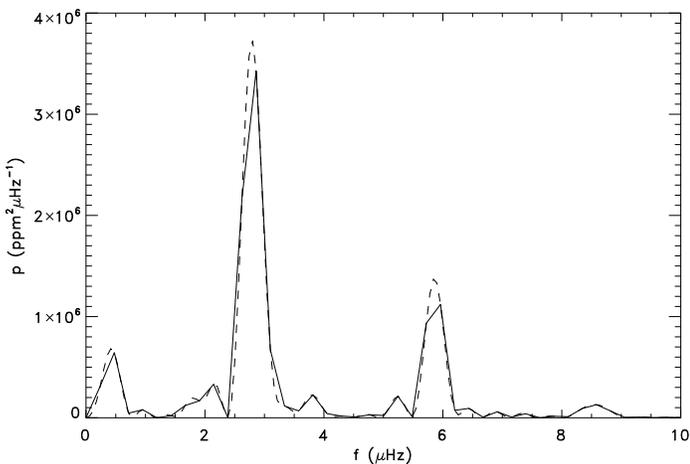}
\vskip 0.1cm
\caption{Zoom on the spectrum at low frequency (dashed line: oversampled fast Fourier transform). The signature of the rotation peaks at 2.80\muHz.
\label{bf-spectrum}}
\end{figure}

\begin{figure}
\centering
\includegraphics[width=8.5cm]{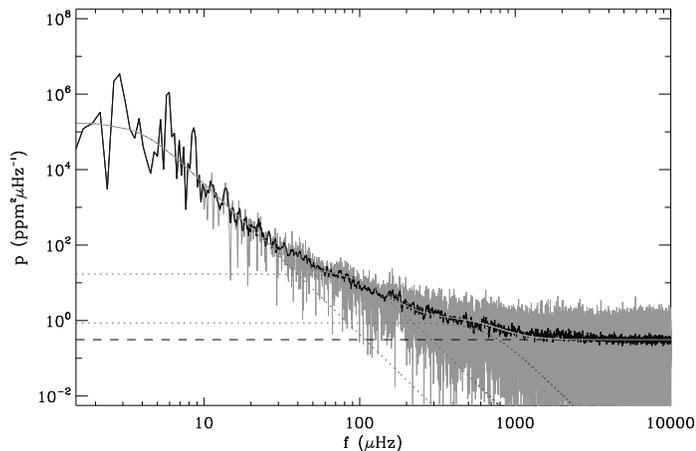}
\caption{Power density spectrum on log-scale axes.
The peaks around 2.8 \muHz\ and its harmonics are signatures of the stellar rotation (cf.\ Fig.~\ref{bf-spectrum}).
The black curve corresponds to a box-car averaged spectrum, with a varying smoothing window (the size of the window increases linearly with frequency).
\label{logscale}}
\end{figure}

\section{CoRoT observations\label{observations}}

This first short run lasted 27.2 days in October 2007 with \cible\ as principal solar-like target. The mean stellar flux, derived from aperture photometry, corresponds to $4.36\ 10^6$ electrons per unit integration. At the usual 32-s sampling of \corot\ seismic data, the mean total flux in the 73\,432 data points is about $1.11\ 10^8$ photoelectrons (Fig.~\ref{intensity}).

\subsection{Gaps and interpolation}

According to the status flag of the data delivered by the chain analysis, 10.3\,\% of the data were lost or affected by some suspect event, mainly due to the South Atlantic Anomaly (SAA) crossing where high-energy particles preclude precise photometric observation. The gaps are responsible for the duty cycle of about 89.7\,\%, with 65\,887 original data points, and the remainders obtained from interpolation. The signature of the orbital period seen in the power spectrum of the window is not purely harmonic, since the repetition of the lost data is modulated by the changing location and size of the SAA area (Fig.~\ref{specfen}). We retrieve the complex pattern reported by \cite{2009arXiv0901.2206A}, with sidelobes at the CoRoT orbital frequency and its harmonics and additional  modulation due to the diurnal frequency. Power in the main sidelobes represents about 1\,\% of the total power, as expected from the observed duty cycle.

Due to the large low-frequency variations in the time series possibly combined with the incomplete window, we have to take care of the quality of the interpolation used for replacing the missing or possibly corrupted data, in order to avoid power leaking to high frequency.
Therefore, we have replaced the contaminated data by values derived from the low-frequency interpolation of the time series based on the algorithm proposed by \cite{atrou} (Fig.~\ref{interpol}). This treatment proved to be efficient, as shown in Fig.~\ref{specbf}, where we verify that the power at medium frequency, where granulation is suspected, cannot be due to the power leaking from the low frequency peaks.

The low-frequency interpolation does not correct the power residuals at the high harmonics of the orbital frequency. These harmonics, due to the window and the high-frequency signal where the seismic signature is searched for, still dominate the spectrum up to 2\,mHz. Therefore, we had to correct the power observed around the harmonics of the satellite rotation, by replacing the spurious peaks by the local mean power observed in the spectrum.

\subsection{Time series and low-frequency pattern}

The time series shown in Fig.~\ref{intensity} exhibits a strong periodic modulation, with a period of about 4 days and a peak to peak amplitude of about 1\,\%. A detailed analysis of this time series is given by \cite{mosser2009}. In the Fourier spectrum, this modulation results in a strong peak at about 2.8\muHz, and corresponds to a rotation period of about 4.0 days. (Fig.~\ref{bf-spectrum}).  According to this measurement and to the \vsini\ value, the star seems to be seen nearly edge-on.

Following \cite{2009A&A...495..979M}, we propose a fit for the stellar background component in the low-frequency pattern, with three Lorentzian-like components in the low-frequency range (below 1\mHz):
\begin{equation}
P(\nu) = \sum_{i=1}^3 {A_i\over 1 + \left(\displaystyle{\nu\over \nu_i}\right)^4}.
\label{lowfreqfit}
\end{equation}
As in \cite{2009A&A...495..979M}, we note that an exponent of 4 provides a better fit. Values of the parameters are given in Table~\ref{low-freq}. An exponent of 2, as in the original model proposed by \cite{1985shpp.rept..199H}, seems  to be excluded by the mean slope observed in the frequency range [4, 40\muHz].  Following \cite{baudin2009}, the first signature, with a characteristic frequency at about 4\muHz\ ($P\sim3$ days), is related to the lifetimes of the activity structures modulated by the stellar rotation. The signatures at higher frequencies are related to the different scales of granulation, similarly to what is observed in photometry in the solar case (\cite{2002ESASP.508...47A}).

\begin{table}
\caption{Parameters of the low-frequency fit.}\label{low-freq}
\begin{tabular}{rllll}
\hline
$\nu_i$ (\muHz)      & 3.96        &  106 & 671 \\
$A_i$ (ppm$^2$/\muHz)& 1.73 10$^5$ &   17 & 0.86  \\
\hline
\end{tabular}\end{table}

\begin{table*}
\caption{Mean power densities}\label{contrib}
\begin{tabular}{lcccccc}
\hline
      & $f\ind{max}$ &\multicolumn{4}{c}{mean power density at maximum (ppm$^2$\muHz$^{-1}$) } & \\
star  & (mHz)        & total & granulation & noise & p~modes   & $\langle$SNR$\rangle$ \\
\hline
HD 49933 & 1.8 & 0.44 & 0.11 & 0.14 & 0.19 & 1.18 \\
HD 181420& 1.6 & 0.65 & 0.20 & 0.28 & 0.17 & 0.79 \\
HD 181906& 1.8 & 1.01 & 0.10 & 0.79 & 0.12 & 0.39 \\
\hline
HD 175726& 2.0 & 0.41 & 0.05 & 0.31 & 0.05 & 0.37 \\
\hline
\end{tabular}

Contributions to the power density at the frequency where the seismic signal is maximum, and mean seismic signal-to-noise ratio for \cible, compared to other \corot\ targets.
\end{table*}

\subsection{High-frequency pattern}

The high-frequency variations of the time series, after high-pass filtering above 0.8\,mHz, present a standard deviation of about 100\,ppm, in agreement with the 95\,ppm value expected from pure photon noise for such a star. Photon-noise limited performance should give a power density of about 0.29 ppm$^2$\muHz$^{-1}$. The observed value is 0.31 ppm$^2$\muHz$^{-1}$.

\begin{figure}
\centering
\includegraphics[width=8.5cm]{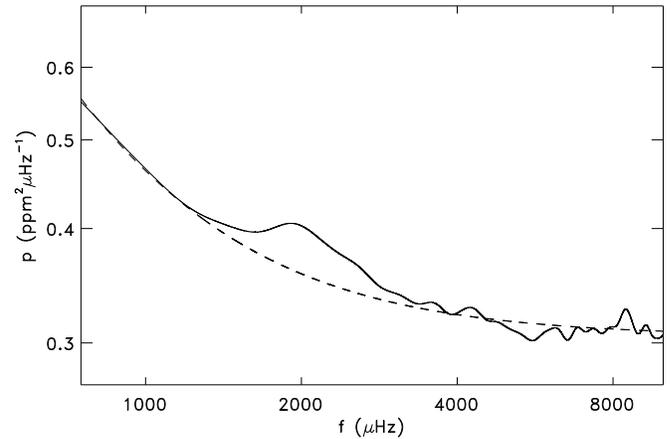}
\caption{Smoothed power spectrum (with an apodized 200-$\mu$Hz FWHM window), on a log-scale axis.
\label{smoothspectrum}}
\end{figure}

\begin{figure}
\centering
\includegraphics[width=8.5cm]{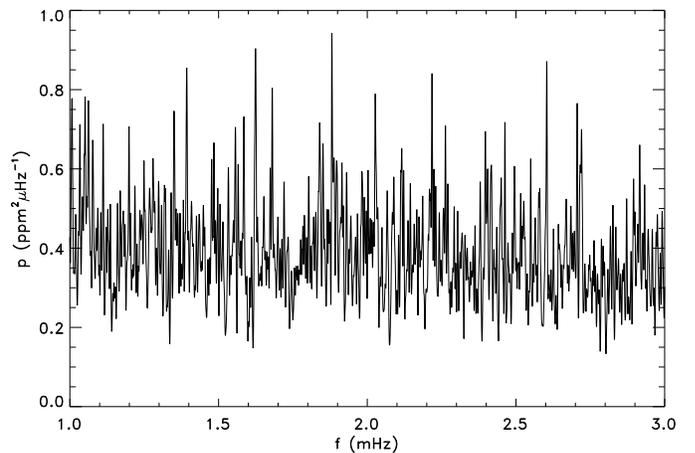}
\caption{Smoothed power density spectrum (with a 2-$\mu$Hz window), around 2\,mHz.
\label{spectre}}
\end{figure}

Contrary to other \corot\ targets showing solar-like oscillations (\cite{2008A&A...488..705A}, \cite{barban2009}, \cite{garcia2009}), no clear signature of pressure modes can be seen in the log-scale plot of the Fourier spectrum (Fig.~\ref{logscale}), even after a slight smoothing process intended to enhance the signal. A strong smoothing of the spectrum is necessary to show evidence of excess power around 2\mHz\ (Fig.~\ref{smoothspectrum}). This signature, even if tiny, corresponds undoubtedly to an  excess of energy, and therefore cannot be confused with a low frequency contribution as described by Eq.~\ref{lowfreqfit}. We notice that this excess of energy appears at a somewhat lower frequency than expected. A plot of the Fourier spectrum centered at 2\,mHz does not exhibit the usual comb pattern expected for pressure modes (Fig.~\ref{spectre}).

From Fig.~\ref{smoothspectrum}, we can derive that the contribution of the seismic power density at 2\,mHz represents only 13\,\% of the total energy density. Table~\ref{contrib} gives the different contributions to the signal for \cible, compared to other targets.  It also gives the mean seismic signal-to-noise ratio (SNR), defined as the square root of the ratio of the smoothed power density of the maximum seismic signal compared to that of the noise.  \cible\ shows the lowest signal-to-noise ratio among the observed asteroseismic targets. Also, the granulation power density is smaller than for other solar-like stars.

\begin{figure}
\centering
\includegraphics[width=8.5cm]{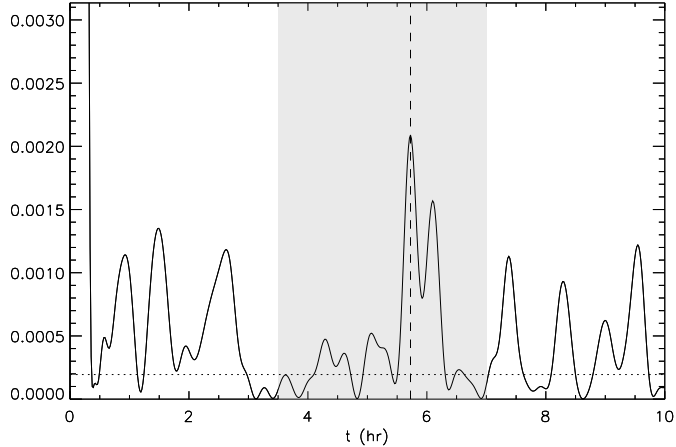}
\caption{Autocorrelation of the time series, calculated as the spectrum of the spectrum filtered with a cosine window centered on 2.0\mHz\ and with 0.8\mHz\ half-width, normalized to 1 at maximum. The dotted horizontal line indicates the mean noise level, derived from the autocorrelation of a pure noise spectrum. The grey region delimits the frequency range where the mean large separation is searched for.
\label{varidnuzero}}
\end{figure}

\section{Data analysis\label{signature}}

\subsection{Determination of the large separation}

A number of different methods were used to analyse the oscillation spectra in the region where power is observed in excess. For the determination of the large separation, we applied an \'echelle diagram analysis, collapsed \'echelle diagrams (\cite{2006ESASP1306..429A}), time frequency analysis, and a test of the H0 hypothesis (\cite{2004A&A...428.1039A}). All methods converge on the same value for the large separation, at about 97\muHz. However, due to the low signal-to-noise ratio, none of the methods were able to give a clear and unambiguous result. For example, systematic trials of collapsed \'echelle diagrams show many possible solutions for a plausible large separation in the range [80--160\muHz]. Also, the \'echelle diagram representations folded on the suspected values are unable to unambiguously confirm any of those values.

The most explicit signature of the large separation is derived from the autocorrelation analysis proposed by \cite{2006MNRAS.369.1491R}. The Wiener-Khinchin theorem relates the autocorrelation function to the power spectral density. Here, we calculate the envelope of the autocorrelation, by determining the power spectrum of the power spectrum, with the Fourier spectrum first filtered by a cosine window covering a frequency range $\nucen \pm \window$, as proposed by \cite{roxburgh2009}
($\nucen$ is the centre of the window and $\window$ its width). The first large peak in the autocorrelation appears at time $\tauto$, related to the local large separation by:
\begin{equation}\label{tauto}
\tauto = {2\over \Delta\nu}.
\end{equation}
This relates to twice the stellar acoustic diameter. When calculated over a large frequency range centered on 2\mHz, the autocorrelation time peaks at about 5.72\,hr, which gives $\Delta\nu\simeq 97.2$\muHz\ (Fig.~\ref{varidnuzero}). Simulations show that this measurement is reliable, despite the noisy aspect of the autocorrelation function.  Calculated with a narrower window (0.35\,mHz), the signature around 5.7\,hr is still present in the autocorrelation of the spectrum windowed from 1.4 to 2.5\,mHz. This would not be the case if the autocorrelation signal was dominated by noise.

One of us (TA) determined that the statistics of the envelope of the autocorrelation signal is a $\chi^2$ with 2 degrees of freedom. Accordingly, we tested the reliability of the signature given by Fig.~\ref{varidnuzero}:
the H0 test (assuming a pure noise contribution) has been rejected at a value of 1\,\%.

\begin{figure}
\centering
\includegraphics[width=8.5cm]{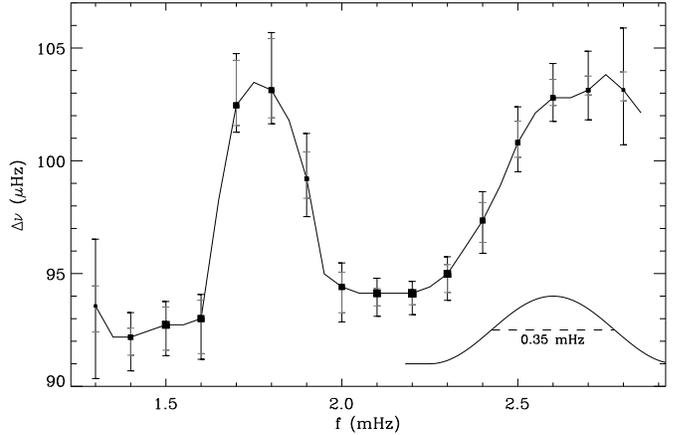}
\caption{Variation of the large separation with frequency, derived from the autocorrelation. The power spectrum was first windowed with a 0.35\mHz-half width cosine filter, as indicated in the lower-right corner. The sampling of the curve corresponds crudely to the mean large separation.
The size of the dots is representative of the amplitude of the correlated signal. Error bars derived from simulations are overplotted: total error is given in black, whereas the grey bar corresponds to the contribution of the internal error.
\label{varidnu}}
\end{figure}

\begin{figure}
\centering
\includegraphics[width=8.5cm]{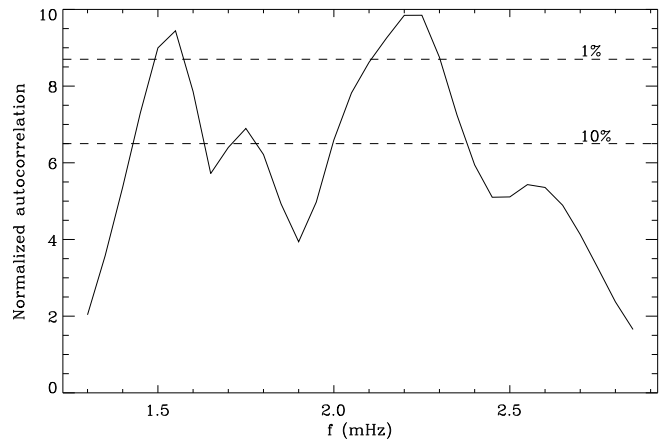}
\caption{Autocorrelation signal as a function of frequency, normalized to the mean signal due to a pure noise spectrum. Dashed lines indicate the level at which the H0 test is rejected.
\label{testvaridnu}}
\end{figure}

\subsection{Variations of the large separation}

\cite{roxburgh2009} describes a method for inferring the variation of the large separations $\deltanunu$ with frequency from the autocorrelation power by taking narrow windows centered on different frequencies.  We have applied this procedure, taking windows centered on different frequencies $\nucen$ in the range [1.5 - 2.5\mHz], with
narrow windows $\window$ varying from 0.2 up to 0.7\mHz. The maximum peak identified near 5.7\,hr gives then a local
value of $\deltanunu$ (Fig.~\ref{varidnuzero}). To test this technique and determine realistic error bars, we constructed artificial power spectra with varying large separations, and repeated the same analysis on both noise-free power spectra and ones with noise. To demonstrate its reliability, we also performed H0 tests.

Thanks to the synthetic power spectra, we identified three main contributions to error bars: time resolution, internal error, and noise, which we will discuss here:\\
- The relative uncertainty on the large separation due to the time resolution is defined by $2 \delta t / \tauto$, with $\delta t$ the sampling time (32\,s). The absolute value, $\delta t\, \langle\Delta\nu\rangle^2$, is of about 0.30\muHz. This incompressible term represents the smallest contribution to the error bars for this noisy target.\\
- The internal error of the method is due to the cosine window. This window allows us to select a given band-pass in the Fourier spectrum, but simultaneously, it correlates the resulting autocorrelation signal for all frequencies in this bandpass. Therefore, the internal error increases with increasing $\window$ value, since a large value of $\window$ smooths out any variation of the large separation $\deltanunu$; simultaneously, the internal error decreases when $\deltanunu$ is nearly constant. For the same reason, this error is not symmetrical. Simulations help us to estimate the asymmetric error bars.
The contribution of the internal error is shown in grey in Fig.~\ref{varidnu}.\\
- The third term contributing to the error bar is related to the interference between noise and signal. The noise complicates significantly the autocorrelation function as shown by Fig.~\ref{varidnu}, and compared to solar results obtained by \cite{2006MNRAS.369.1491R}. At low seismic amplitude, this term represents of course the main source of error. At maximum signal, around 2\,mHz, the autocorrelation peaks have a signal-to-noise ratio of about 10.

Simulations also show that the measured variation of the large separation corresponds to a global change of the large separations of modes $\ell=0$ and 1, and is not sensitive to a degree-dependent modulation as should be produced by a structure discontinuity (\cite{1993A&A...274..595P}). We verified that the local increase of the large separation in the range [1.6 - 1.9\,mHz] is not due to a possible misidentification of ridges, a common artifact in noisy or single-data seismic data (e.g. \cite{2008A&A...478..197M}).

We also tested the reliability of the pattern given by Fig.~\ref{varidnu} with the H0 test. The level at which the test is rejected is indicated in Fig.~\ref{testvaridnu}. It shows that the bump in the frequency range [1.65 - 1.9\,mHz] is less confident than the lower values in the surrounding frequency ranges. If real, the modulation of the large separation with a period $W$ of the  order of 0.8--1.0\mHz\ could  be caused primarily by the He\,II ionisation zone or be due to the internal phase shifts. Following \cite{2005MNRAS.361.1187M}, we infer the acoustic depth $\tau$ of this region which is $\tau \simeq 1/2W$ resulting in the range $\tau=500$--$670$\,s.

\begin{figure}
\centering
\includegraphics[width=8.5cm]{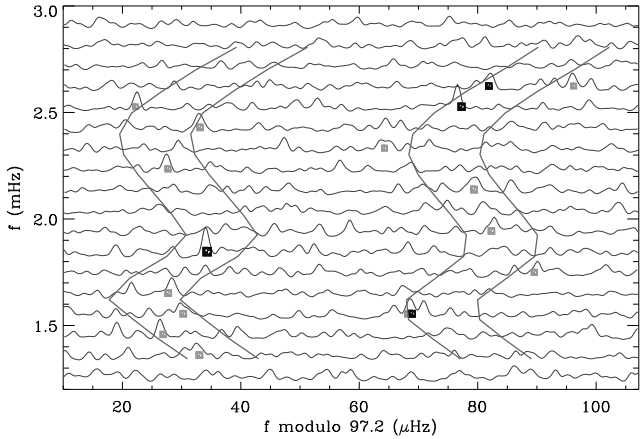}
\caption{\'Echelle diagram of the smoothed spectrum, with indication of the possible ridges identified  according to the varying large separations shown in Fig. \ref{varidnu}. The grey squares indicate the location of the peaks with a height-to-background ratio greater than 3; the black squares indicate the location of the most confident peaks (above 95\,\%) identified by the H0 test.
\label{visimod}}
\end{figure}

\begin{figure}
\centering
\includegraphics[width=8.5cm]{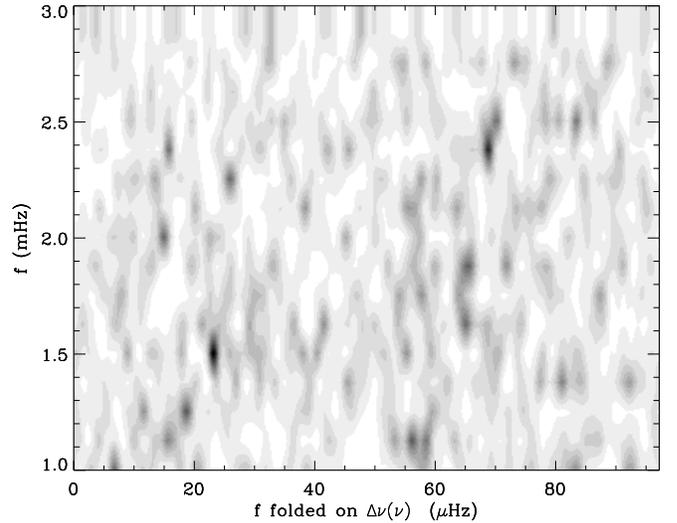}
\caption{Rectified \'echelle diagram, folded according to the function $\deltanunu$ plotted in Fig.~\ref{visimod}. The major part of the energy is now close to two vertically aligned ridges around 20 and 70\muHz.
\label{rectified}}
\end{figure}

\begin{figure}
\centering
\includegraphics[width=8.5cm]{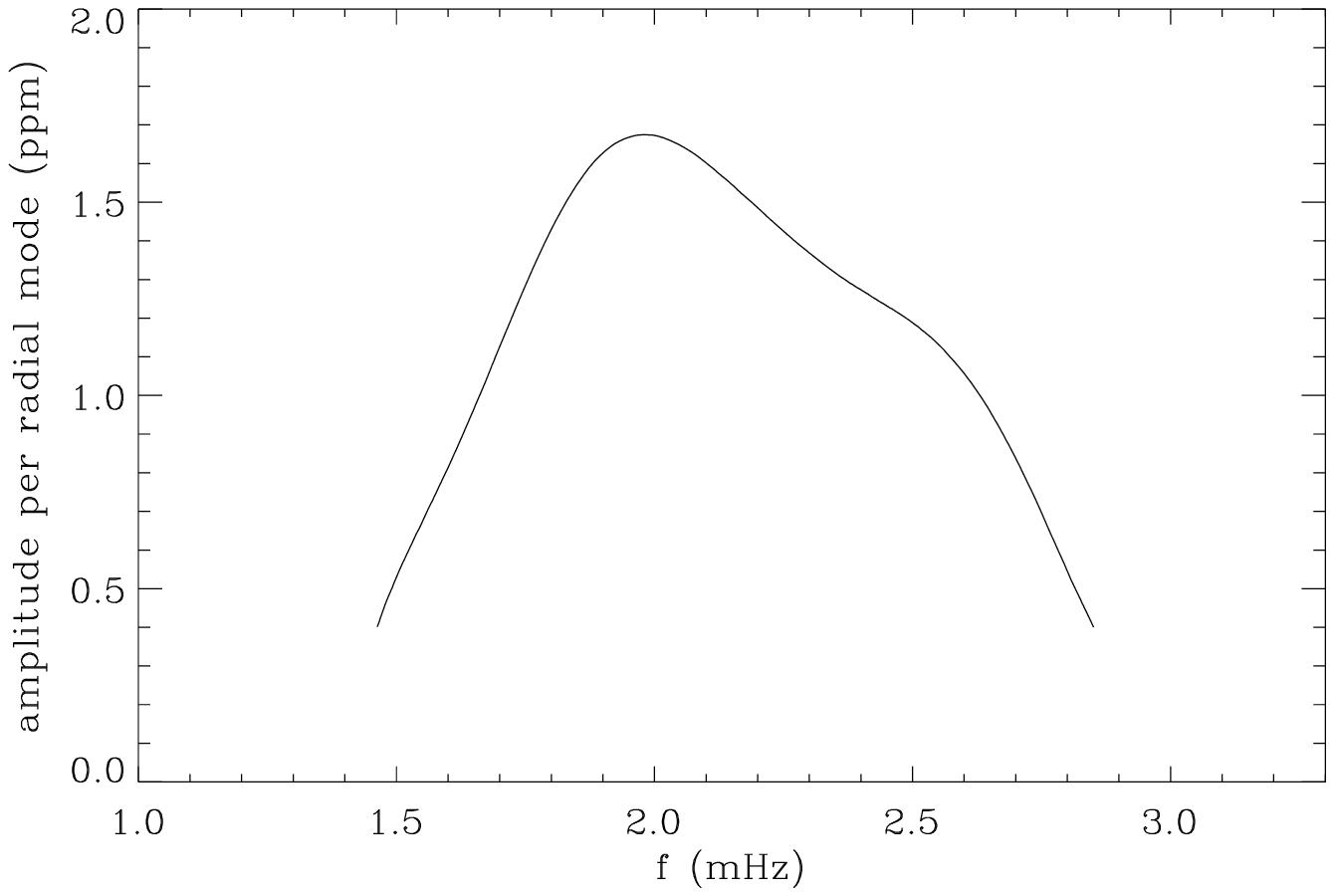}
\caption{Bolometric amplitude per radial mode, calculated according to \cite{{2009A&A...495..979M}}.
\label{ampli}}
\end{figure}

\subsection{Analysis of the frequency pattern}

Given the estimates of the function $\deltanunu$ and assuming it is real, we have tried to use this information to extract eigenfrequencies. First, we have searched for the peaks that show the highest height-to-background ratios (Fig.~\ref{visimod}). Most of the major peaks are located in 2 ridges, presumably related to modes with even and odd degrees. We then constructed a rectified \'echelle diagram, where the spectrum is not folded on a fixed frequency, but on varying frequencies according to the function $\deltanunu$. This representation then helps to identify vertically aligned ridges (Fig.~\ref{rectified}). We applied the H0 test to the selected peaks: only 4 out of the peaks have less than 10\,\% probability of being due to pure noise. The frequencies of these peaks are indicated in Table~\ref{identi}.

\begin{table}
\caption{Highest peaks}\label{identi}
\begin{tabular}{cccc}
\hline
order& $\Delta\nu$ & $\nu_1$ & $\nu_2$ \\
     &\multicolumn{3}{c}{\dotfill\ ($\mu$Hz) \dotfill}\\
\hline
 13 &   92.1 &  1393.7 &           \\
 14 &   92.6 &  1485.0 &           \\
 15 &   92.9 &  1585.7 &  \textit{1623.3}   \\
 16 &  100.4 &  1680.4 &           \\
 17 &  103.2 &         &  1839.2    \\
 18 &  100.2 &  \textit{1881.1}           \\
 19 &   94.6 &         &  2027.6   \\
 20 &   94.1 &         &           \\
 21 &   94.1 &         &  2217.3   \\
 22 &   94.5 &  2263.1 &           \\
 23 &   96.3 &         &  2397.3   \\
 24 &   99.0 &  2463.1 &           \\
 25 &  102.1 &  2550.0 &  \textit{2604.2} \\
 26 &  102.8 &         &  \textit{2706.3}  \\
 27 &  103.7 &         &           \\
 28 &  101.9 &         &           \\
\hline
\end{tabular}

Variation of the large separation along the frequency pattern, and largest peaks corresponding to the 2 ridges; the  peaks selected with the H0 test at a confidence level greater than 95\,\% appear in italic.
\end{table}

The surface rotation period being determined from the light curve, we have searched for the signature of rotational multiplets, without any success. A precise analysis of these ridges is out of the scope of this paper. Due to the low signal-to-noise ratio, no precise eigenmode identification is expected.

\subsection{Mode amplitudes and lifetimes\label{amplit}}

Due to the impossibility of reliably extracting many individual peak, mode amplitudes were determined according to the global recipe exposed in \cite{2009A&A...495..979M}. According to this work, the maximum bolometric amplitude per radial mode is about 1.7$\pm$0.25\,ppm (Fig.~\ref{ampli}).

With such a low signal-to-noise ratio, estimating lifetimes in the spectrum is at least as difficult as estimating eigenfrequencies. A clue indicating  that the lifetimes are short is given by the use of the H0 hypothesis (\cite{2004A&A...428.1039A}). Appourchaux addresses the case of the H0 test with binned data, in order to enhance the detection of modes with finite lifetime. Most of the peaks revealed by the analysis above remain undetectable with the test applied to unbinned data, but can be found when considering large bins. The maximum number of peaks exceeding the limit defined by the H0 test occurs as soon as 9 bins are considered, which corresponds to 3.8\muHz. Accordingly, the  mean mode lifetime is then estimated to be about 3 days.

\section{Discussion\label{discussion}}

\subsection{The small observed value of the large separation}

The observed value of the mean large separation at 97.2$\pm$0.5\muHz\ is significantly lower than the expected value (132$\pm$20\muHz), but is fully compatible with the value already noticed for the frequency of maximum power. We consider that the unambiguous excess power at 2\,mHz (Fig.~\ref{smoothspectrum}) reinforces the identification of the large separation.
The observed large separation and in particular the frequency of maximum power indicate a less dense star than our target, which is a quasi-solar twin as deduced from the fundamental parameters.
In order to reconcile these seismic values with the fundamental parameters, we have to examine in which cases these parameters could be incorrect, or translate into incorrect mass and radius. We are aware that this requires a change of about 3.5\,$\sigma$ compared to the nominal values of the mass and radius, respectively around $0.85\, M_\odot$ and $1.15\, R_\odot$. For comparison, in Table~\ref{prop-phys}, we present the parameters $1\, M_\odot$ and $1\, R_\odot$ from \cite{bruntt2009}.

An error in the luminosity of \cible\ could significantly modify the inferred seismic parameters. We first verified that the photon count registered for \cible\ is in agreement with its V magnitude, by comparison with other targets. Contrary to the case of HD\,181906, \cite{bruntt2009} does not suspect \cible\ of being member of a double system according to the spectrum analysis. The SIMBAD database identifies it as the 3rd component of the system $\theta$ Serpentis, but 7 arcmin away from the two major components. We further checked in the CoRoT database that the point spread function of \cible\ is nominal. This means that a second component, if present, should have a magnitude fainter than 13.5 or be very close to \cible. Since a 6-magnitude fainter star would not affect the determination of the fundamental parameters, we consider more probable the possible presence of a very close object with a negligible spectral signature (for instance due to very broad lines). However, this explanation fails to explain the discrepancy: a lower flux for \cible\ yields a lower mass, hence a higher density and a higher large separation.

An error on the mean effective temperature is another possible explanation, for example due to a rapidly rotating star seen nearly pole-on. However, this seems to be excluded by the spot analysis achieved by \cite{mosser2009}, in favor of an inclination in the range [60 - 90$^\circ$] when taking into account all available information on this star.

The large discrepancies reported by the different works addressing the fundamental parameters of \cible, as reported in Section~\ref{etoile} and in \cite{bruntt2009}, should favor the hypothesis that some unexplained feature affects the high-resolution spectrum and perturbs its analysis.
The pressure sensitive calcium lines at 6122 and 6162\,\AA\ show for instance that either $\log g > 4.9$ or that the Ca abundance in these lines is slightly higher than the nearly solar abundance measured from several weak Ca lines. This discrepancy indicates that the spectrum of \cible\ is not completely understood.

At this stage, we fail to understand the observed discrepancies. Further modeling of the stellar spectrum is necessary. In forthcoming work we will present an interior structure model derived from the observed seismic parameters. Also, we will acquire high angular resolution imaging with adaptive optics, to examine any possible faint stars in the vicinity of the star.

\subsection{The low observed mode amplitudes}

The amplitude reported in Section~\ref{amplit}  is 1.7 times less than expected, significantly below the value derived from the scaling law. We already noticed that undermetallic stars show low amplitude (\cite{2008A&A...478..197M}). \cite{samadi2009} have recently quantified the effect of the surface metal abundance on the efficiency of the mode driving excitation by turbulent convection.  However, \cible\ is only slightly undermetallic, and other reasons are needed to explain the low observed amplitudes.

The low amplitudes are likely related to the large activity signal of \cible. This observation reinforces the idea that a large magnetic field inhibits the stellar convection, hence the p~mode excitation, as was shown for the Sun by \cite{2008ApJ...684L..51J}. Part of the discrepancy may also be due to an overestimated luminosity, but this seems to be excluded from the discussion above.

\section{Conclusion\label{conclusion}}

The analysis of the spectrum of \cible\ is hampered by the very low seismic signal and a complicated spectrum showing large discrepancies compared to the 2nd order Tassoul-like oscillation pattern. At signal maximum, the mean signal-to-noise ratio (derived from the square root of the smoothed power density) is only 0.37. Despite the fact that the granulation signal is smaller compared to other CoRoT targets, as given in \cite{2008Sci...322..558M}, the power density of the seismic signal represents only 13\,\% of the total power.

To best extract information from the power spectrum of \cible, we have determined the most efficient methods. The autocorrelation method proves to give the best result in extracting the large separation. Furthermore, even with a low signal-to-noise ratio, it allows us to extract further information on the variation of the large separation with frequency. This shows an increase of the large separation with frequency and a possible modulation. The H0 test confirms this modulation only at a level of about 85\,\%. If real, this modulation is compatible with the seismic signature of the He\,II ionization zone. Information about small separations cannot be derived due to the weak signal.

As already discussed by \cite{2006MNRAS.369.1491R}, the method based on the autocorrelation power to extract information from a noisy spectrum is very cheap computationally, and can be particularly productive in some very common cases for the \corot, Kepler and Plato missions: efficient initial analysis of the data and in particular the analysis of noisy data.

The large separation and location of the maximum signal  we can derive for this star disagree with the expected values and this will require further analysis.
The maximum amplitude shows a strong deficiency compared to the value expected from the scaling law in $(L/M)^{0.7}$. This reinforces the idea that a large magnetic field inhibits the stellar convection, hence the p~mode excitation.

Comparison with other \corot\ targets proves that the short-run data set is prejudicial for retrieving precise eigenfrequencies in the oscillation spectrum. We have compared the spectrum of \cible\ with the ones of \corot\ solar-like targets, calculated for a similar reduced 27.2-day time span.
According to their greater mean power densities, the spectra of the short time series of HD\,49933 (\cite{2008A&A...488..705A}) and HD\,181420 (\cite{barban2009}) still show a clear seismic signature. However, the spectrum of HD\,181906 (\cite{garcia2009}), when reduced to 27.2 days, shows no clear signal, contrary to the full time series spectrum. Since \cible\ and HD\,181906 have very similar signal-to-noise ratios, we can expect that mode identification of \cible\ will be possible in a long-run spectrum. \cible\ presents the lowest effective temperature among current solar-like \corot\ targets. A long run would permit us to extend the CoRoT seismic analysis to a much wider region of the HR diagram.

In most aspects (mass, radius, chemical composition), \cible\ seems to be a solar twin, differing by a somewhat higher effective temperature (300~K) and luminosity (21\%), and a slightly lower metallicity. On the other hand, the asteroseismic parameters indicate a significantly less dense star; this discrepancy reinforces the importance of continuing studies of the star.

\begin{acknowledgements}
JB acknowledges support through the ANR project Siroco.
\end{acknowledgements}

\end{document}